\documentclass[superscriptaddress,onecolumn]{revtex4-1}
\pdfoutput=1
\usepackage{amsmath,bm,epsfig,epstopdf}
\usepackage[]{graphicx}
\usepackage{subfig}
\usepackage{color}
\begin{document}
\title{Where do rivers grow? Path selection and growth in a harmonic field}
\author{Yossi Cohen}
\affiliation{Lorenz Center, Department of Earth Atmospheric, and Planetary Sciences,
Massachusetts Institute of Technology, Cambridge, MA. 02139, USA}
\author{Olivier Devauchelle}
\affiliation{Institut de Physique du Globe, 4 place Jussieu, 75252 Paris cedex 05, France}
\author{Hansj\"{o}rg F. Seybold}
\author{Robert S. Yi}
\affiliation{Lorenz Center, Department of Earth Atmospheric, and Planetary Sciences,
Massachusetts Institute of Technology, Cambridge, MA. 02139, USA}
\author{Piotr Szymczak}
\affiliation{Institute of Theoretical Physics, Faculty of Physics, Warsaw University, Ho\.{z}a 69, 00-681 Warsaw, Poland}
\author{Daniel H. Rothman}
\affiliation{Lorenz Center, Department of Earth Atmospheric, and Planetary Sciences,
Massachusetts Institute of Technology, Cambridge, MA. 02139, USA}
\date{\today}


\begin{abstract}
River networks exhibit a complex ramified structure that has inspired decades of studies. Yet, an understanding of the propagation of a single stream remains elusive. Here we invoke a criterion for path selection from fracture mechanics and apply it to the growth of streams in a diffusion field. We show that a stream will follow local symmetry in order to maximize the water flux and that its trajectory is defined by the local field in its vicinity. We also study the growth of a real network. We use this principle to construct the history of a network and to find a growth law associated with it. The results show that the deterministic growth of a single channel based on its local environment can be used to characterize the structure of river networks.
\end{abstract}
\maketitle

\section{Introduction}
As water flows it erodes the land and produces a network of streams and tributaries \cite{45H,80D,RR01}. Each stream continues to grow with the removal of more material, and evolves in a direction that corresponds to the water flux entering its head. The prediction of the trajectory of a growing channel and the speed of its growth are important for understanding the evolution of complex patterns of channel networks.
Several models address their evolution and ramified structure. One, the Optimal Channel Networks model \cite{96MCFCB}, is based on the concept of energy minimization and suggests a fractal network. The landscape evolution method and many diffusion-based models \cite{98TB,97FD,09PKD} have also proven useful for modeling erosion and sediment transport. These models distinguish between two regimes: one is a diffusion-dominated regime where topographic perturbations are diminished, which leads to a smoother landscape and uniform symmetric drainage basins. In this case, the shape of a channel cannot deviate from a straight line. In the second regime, advection dominates, and channel incisions are amplified. The channel effectively continues to the next point that attracts the largest drainage basin, which corresponds to the direction where it receives the maximum water flux. These models nicely predict the formation of ridges and valleys in an advection-diffusion field and provide insight into the interaction between advective and diffusive processes \cite{72SB,08PDK}. Yet they do not address directly the evolution of a single channel and do not explicitly address the nature of a growing stream based on its local environment.

Here we address two basic questions in the evolution and the dynamics of a growing channel: where it grows and at what velocity. We propose that the direction of the growth of a stream is defined by the flux field in the vicinity of the channel head. This theory is widely used in the framework of continuum fracture mechanics and accurately predicts crack patterns in different fracture modes, for both harmonic and bi-harmonic fields and for different stress singularities \cite{61BC,74GS,10CP}. The theory, known as the principle of local symmetry, states that a crack propagates along the direction where the stress distribution is symmetric with respect to the crack direction \cite{61BC,80CR}. We find an analog of this principle in the motion and growth of channels in a diffusive field. We argue that a stream will evolve in a direction that maximizes the water flux, and its trajectory is dictated by the symmetry of the field in the vicinity of its head. We demonstrate how to determine a growth law that ties the water flux into a channel to an erosion process and the propagation velocity of a channel head.


\section{Principle of local symmetry}
We study the case of channel growth driven by groundwater seepage as a representative process for channel formation and growth in a diffusing field \cite{69D, 80D, 90D,93DD,09ALPSMMKR,11DPLR,12PDKR}. The emergence of groundwater through the surface leads to erosion and the development of a drainage network~\cite{09ALPSMMKR}.
The flow of groundwater is described by Darcy's law \cite{80D}
\begin{equation}
\mathbf{v}=-\kappa\nabla\left(\frac{p}{\rho g} + z\right).
\end{equation}
Here $\mathbf{v}$ is the fluid velocity, $\kappa$ the hydraulic conductivity, $p$ the pressure in the fluid, $\rho$ the fluid density, $z$ the geometric height, and $g$ the gravitational acceleration.
By assuming only horizontal flow, the Dupuit approximation \cite{B72, D63} relates the water table height $h(x,y)$ to the groundwater horizontal velocity $\mathbf{v}=-\kappa\nabla h$, and to the groundwater flux $\mathbf{q}=-h\kappa\nabla h=-\frac{\kappa}{2}\nabla h^2$.
Considering an incompressible flow, the steady state solution for the water table height becomes a function of the ratio between the mean precipitation rate $P$ and $\kappa$,
\begin{equation}
\frac{\kappa}{2}\nabla^2h^2=-P.
\label{dupuit}
\end{equation}
Thus, the square of the height $h$ is a solution of the Poisson equation \cite{D63,PK62}. By rescaling the field, Eq. \eqref{dupuit} becomes
\begin{equation}
\nabla^2\phi=-1,
\label{rdupuit}
\end{equation}
where $\phi=(\kappa/2P)h^2$, and $-\nabla \phi$ is the Poisson flux. The boundaries are given by the stream network; for a gently sloping stream we can assume that the water table elevation at the boundary is $h=0$, and therefore $\phi=0$ along the streams.

\begin{figure}[h]
  \includegraphics[scale=0.5]{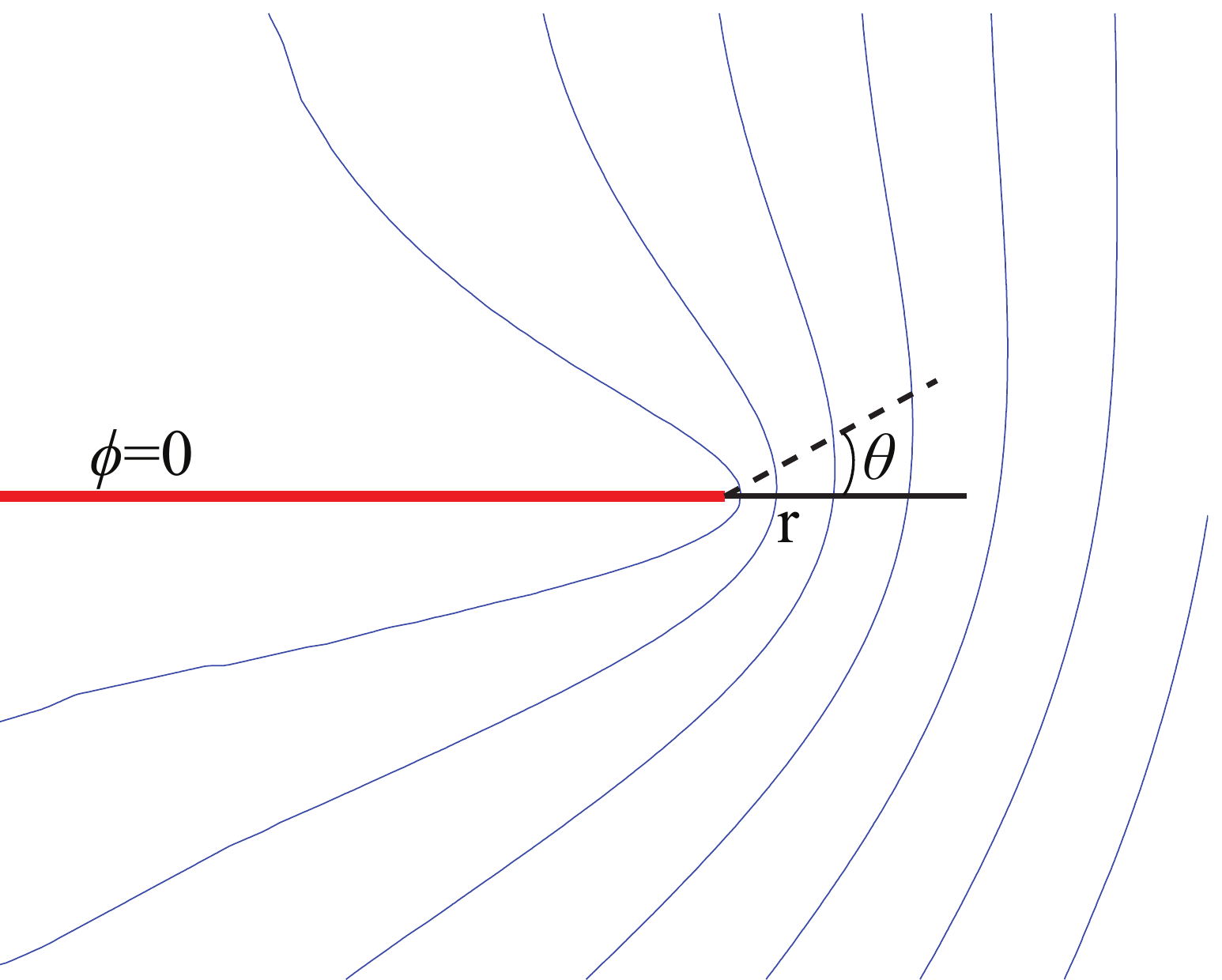}
  \caption{A channel (in red) in a Poisson field. The equipotential lines of the field are in blue. $r$ is the distance from the channel head, and $\theta=0$ indicates the growth direction of the channel.}
  \label{expan}
\end{figure}

The slow erosion of sediment into the stream is an increasing function of the water flux into the channel head. Thus, a spring will grow in a direction in which the groundwater flows. In the vicinity of the channel head we can neglect the Poisson term in Eq. \eqref{rdupuit}, and the field can be approximated as \cite{13PDSR,12DPSR},
\begin{equation}
\nabla^2\phi=0.
\label{laplace}
\end{equation}
For a semi-infinite channel on the the negative $x$-axis with boundary conditions
\begin{equation}
\phi(\theta=\pm\pi)=0,
\end{equation}
the harmonic field around the tip can be expressed in cylindrical coordinate as \cite{13PDSR}
\begin{equation}
\phi(r,\theta)=a_1r^{1/2}\cos\left(\frac{\theta}{2}\right) + a_2r\sin(\theta) + \mathcal{O}(r^{3/2}),
\label{expansion}
\end{equation}
where, as shown in Fig. \ref{expan}, $r$ is the distance from the channel head and the channel is located at $\theta=\pm\pi$. The coefficients $a_i$, $i=1,2$ are determined by the shape of the water flux coming from the outer boundary.

\begin{figure}[h]
\centering
\subfloat[]{
\includegraphics[scale=0.5]{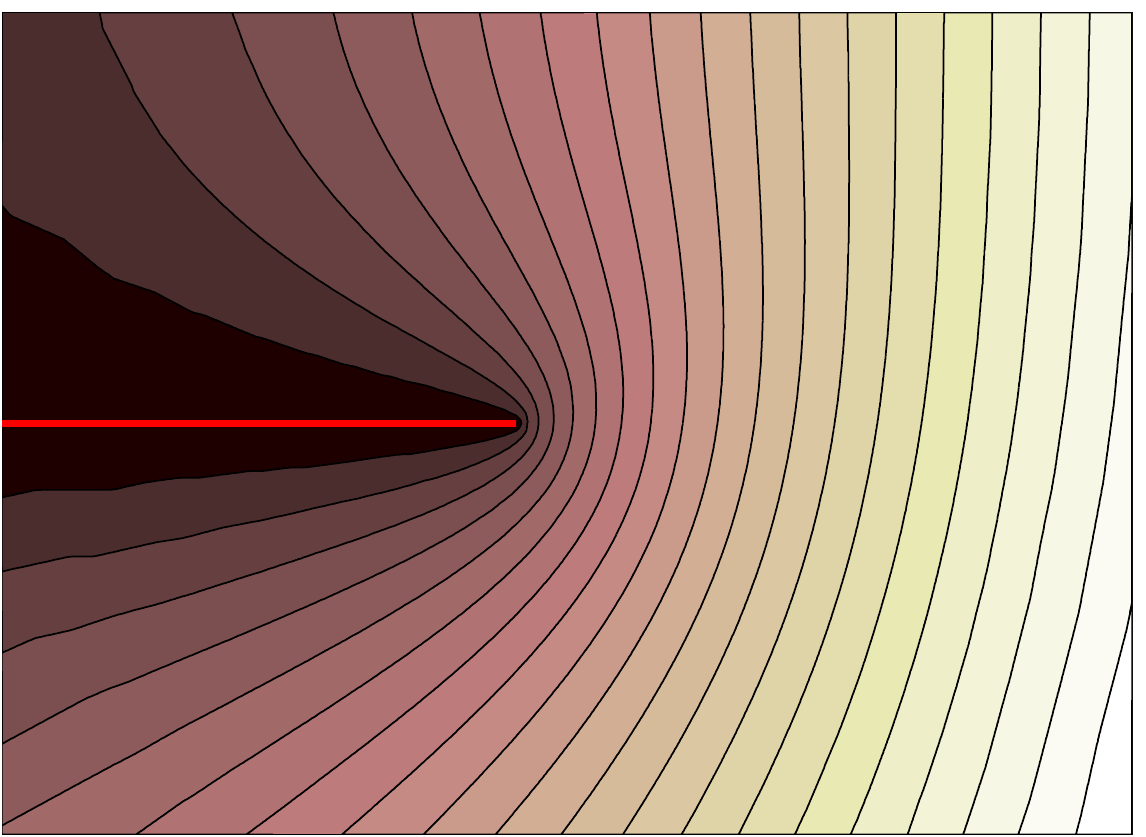}
\label{sc1}
}
\hspace{0mm}
\subfloat[]{
\includegraphics[scale=0.5]{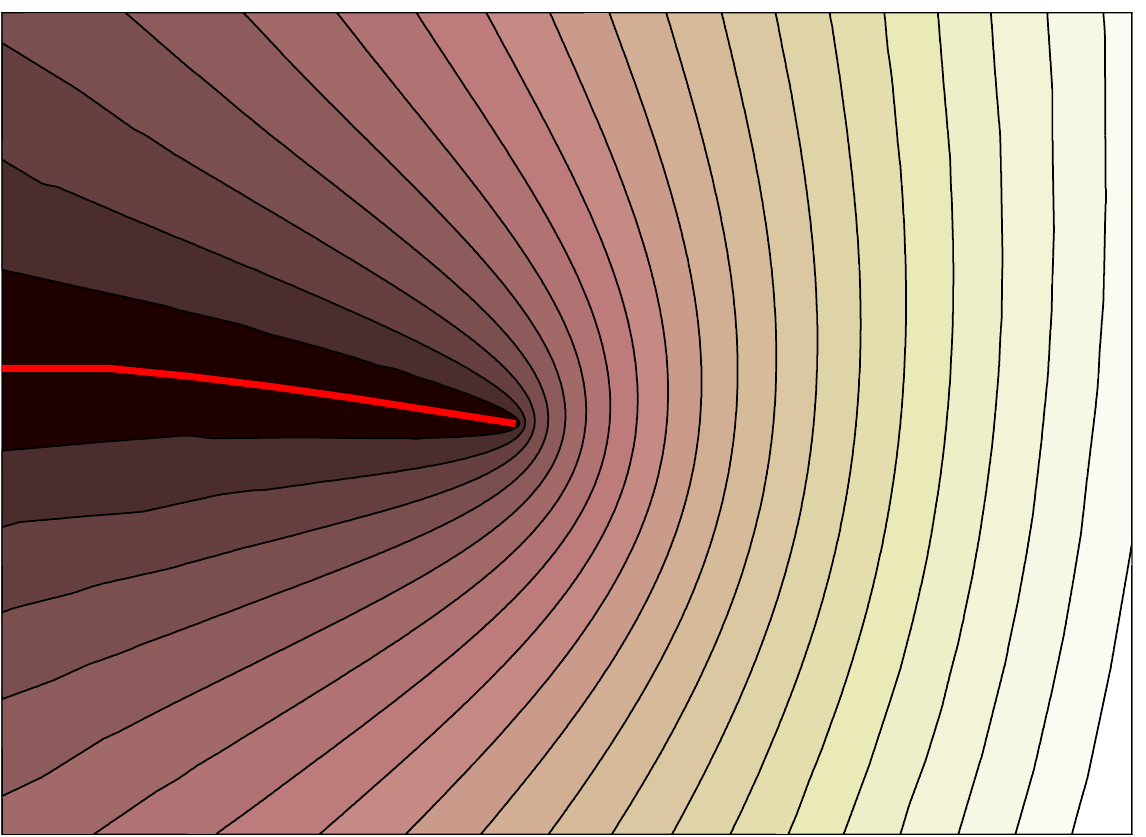}
\label{sc2}
}
\caption{(a.) Asymmetric field ($a_2\neq0$); As the stream grows, it must bend in a direction for which $a_2$ vanishes. (b.) Symmetric field ($a_2=0$): Growth according to the principle of local symmetry.}
\label{sc}
\end{figure}

We hypothesize that a channel evolves in a direction that maximizes the water flux. Since the leading term in the expansion of Eq. \eqref{expansion} is symmetric with respect to $\theta$, it will not reflect any trajectory of the stream other than a straight line. Thus, we must also consider the subdominant term that breaks the symmetry and causes the stream to propagate in a different direction. Other terms in the expansion are negligible in the vicinity of the channel head.
We suggest that this direction is defined by the principle of local symmetry: a stream propagates in the direction for which $a_2$ vanishes. Fig. \ref{sc} expresses this notion pictorially. In fracture mechanics, a crack follows local symmetry in order to release the maximum stress \cite{61BC,80CR}. We therefore propose an analogous criterion for growth of a channel. We proceed to show that this is equivalent to growth in a direction that maximizes flux.


\section{Evaluation of the principle}
We evaluate the principle of local symmetry in two ways: First, we show analytically that growth of a stream in a direction that maximizes the water flux is equivalent to growth that follows local symmetry. Second, we develop a numerical method to propagate streams in complex boundary conditions according to the principle of local symmetry. To validate this method, we consider an alternative model for the growth of a channel in a Laplacian field using the deterministic Loewner equation \cite{02CM,04GK} and we show that the path obtained in this model is consistent with local symmetry.

\subsection{Local symmetry implies maximization of flux}

In the expansion \eqref{expansion} of the field in the vicinity of the channel head, a symmetric field is achieved only when the subdominant term that breaks the symmetry disappears, i.e. when $a_2=0$. Here, we show that seeking local symmetry is equivalent to maximizing the flux.

The water flux as $r\rightarrow0$ corresponds to the gradient of Eq. \eqref{expansion},
\begin{eqnarray}
q_r&=\frac{\partial\phi}{\partial r}&=\frac{a_1}{2r^{1/2}} \cos(\theta/2) + a_2\sin(\theta), \\
q_\theta&=\frac{1}{r}\frac{\partial\phi}{\partial\theta}&=-\frac{a_1}{2r^{1/2}} \sin(\theta/2) + a_2\cos(\theta).
\label{flux}
\end{eqnarray}
The total flux crossing a point close to channel tip can be expressed as
\begin{equation}
\|\nabla\phi\|=\sqrt{q_r^2 + q_\theta^2}=\sqrt{\frac{a_1^2}{4r} + a_2^2 + \frac{a_1a_2}{\sqrt{r}}\sin(\theta/2)}.
\label{totflux}
\end{equation}
Since the water erodes the landscape, the river will curve toward the direction of the maximum flux,
\begin{equation}
\frac{\partial\|\nabla\phi\|}{\partial\theta} = \frac{\frac{a_1a_2}{4\sqrt{r}}\cos(\theta/2)}{\sqrt{\frac{a_1^2}{4r} + a_2^2 + \frac{a_1a_2}{\sqrt{r}}\sin(\theta/2)}} = 0,
\end{equation}
and as $r\rightarrow0$, we obtain
\begin{equation}
a_2\cos(\theta/2)=0.
\end{equation}
Thus, the water flux exhibits extreme values when $\theta=\pm\pi$ or $a_2=0$. For non vanishing $a_2$, the maximum value of the total flux is obtained at one of the channel's sides, and the minimum at the opposite side (Eq. \eqref{totflux}). In this case, the stream will have to curve as the erosion of material is not distributed equally. However, only for the second solution, where the asymmetric term $a_2$ vanishes, the maximum of the flux is obtained in front of the stream and the stream grows in this direction without curving.

\subsection{Laplacian paths maintain local symmetry}

Next, we design a numerical method to calculate trajectories that explicitly maintain local symmetry, and compare its results to an analytic solution.
We consider a simple case in which one channel grows in a confined rectangular geometry ${-1<x<1, 0<y<30}$ in a Laplacian field. We first calculate the trajectory using the principle of local symmetry. Our algorithmic implementation of this principle requires that at each step streams grow in the direction for which $a_2$ vanishes. (Further details are in Appendix A).
We apply the following boundary conditions:  a zero elevation at the bottom ($\phi=0$ at $y=0$), which corresponds to a main river or an estuary; no flux at the sides ($\frac{\partial\phi}{\partial x}=0$ at $x=1,-1$), which corresponds to a groundwater divide; and a constant flux of water from the top, ($\frac{\partial\phi}{\partial y}=1$). We then initiate a small slit ($l=0.01$) perpendicular to the bottom edge, and allow it to grow according to the principle of local symmetry. Not surprisingly, a stream initiated at the middle of the lower edge ($x=0;y=0$) continues straight. However, when we break the symmetry and initiate a slit left of the center ($x=-0.5;y=0$) the stream bends toward the center of the box.

To validate the principle of local symmetry, we compare our numerical trajectory to the evolution of a path in a Laplacian field according to the deterministic Loewner equation \cite{02CM,04GK,08GS}. In the Loewner model, the properties of analytic functions in the complex plane are used to map the geometry into the complex half plane or into radial geometry, and to find the solution for the field. Then, at each time step, a slit is added to the tip of the channel based on the gradient of the field entering the tip.

\begin{figure}[htp]
  \includegraphics[scale=0.3]{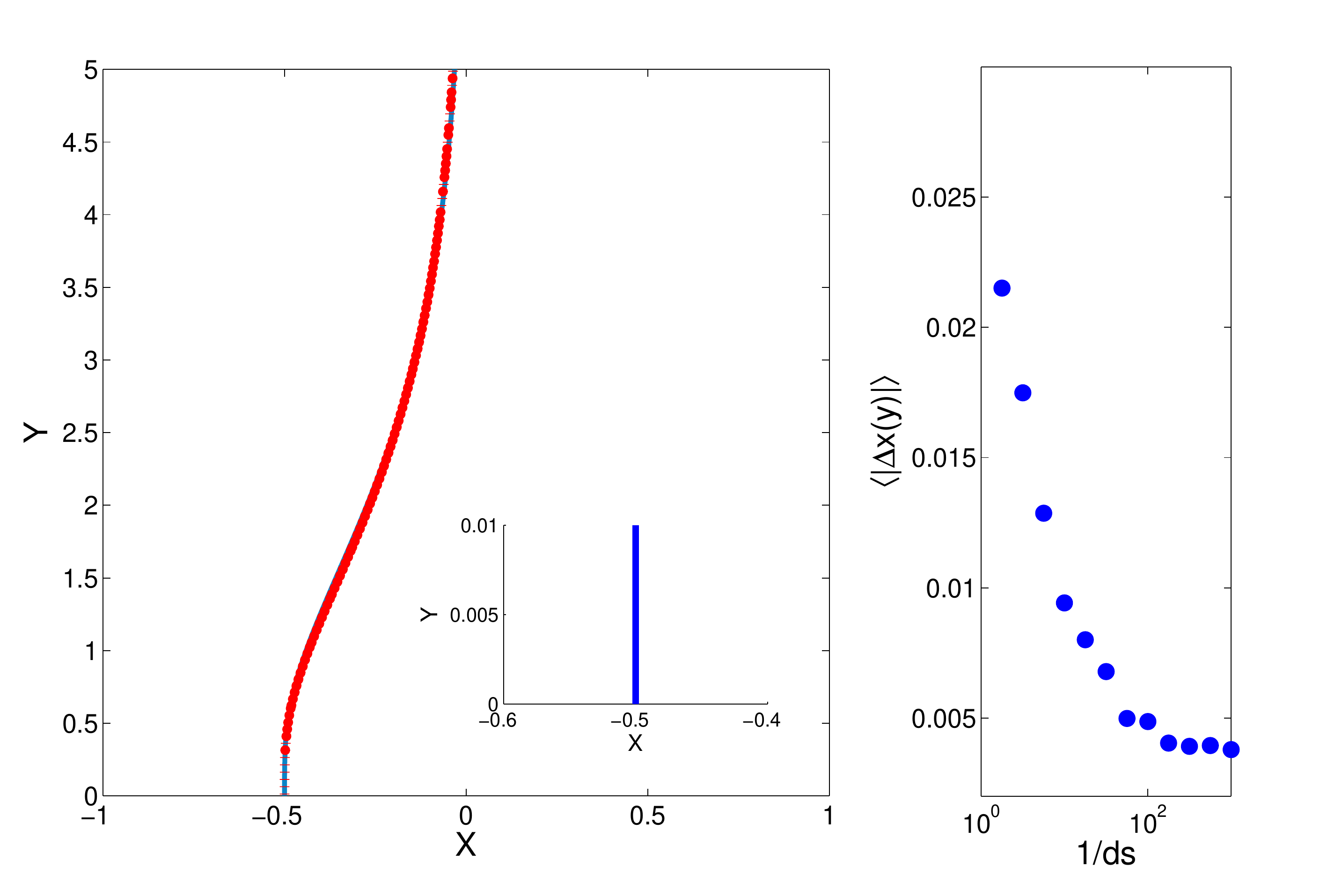}
  \caption{\emph{Left:} Trajectory of a single stream initiated at the bottom, to the left of center. Blue: the analytical solution \cite{08GS}. Red: the numerical trajectory of a stream grown according to the principle of local symmetry. Inset: the initial slit. \emph{Right:} The average error $\langle|\triangle x(y)|\rangle$ between the numerical trajectory and the analytical solution with the decrease of the step size, $ds$.}
  \label{singchannel}
\end{figure}

Fig. \ref{singchannel} compares results from the two approaches. We find that the two solutions exhibit the same trajectory. In Appendix B, we present a proof that the growth of a channel using the Loewner equation always fulfills local symmetry.

\section{Growth of a real stream network.}
The evolution of a channel is defined by the field in the vicinity of the tip. This field is non-local and highly dependent on the ramified network of the streams.
In this section, the numerical method developed in the previous section is used to compute trajectories in more general settings where no analytic solutions are possible.

\begin{figure}
  \includegraphics[scale=0.25]{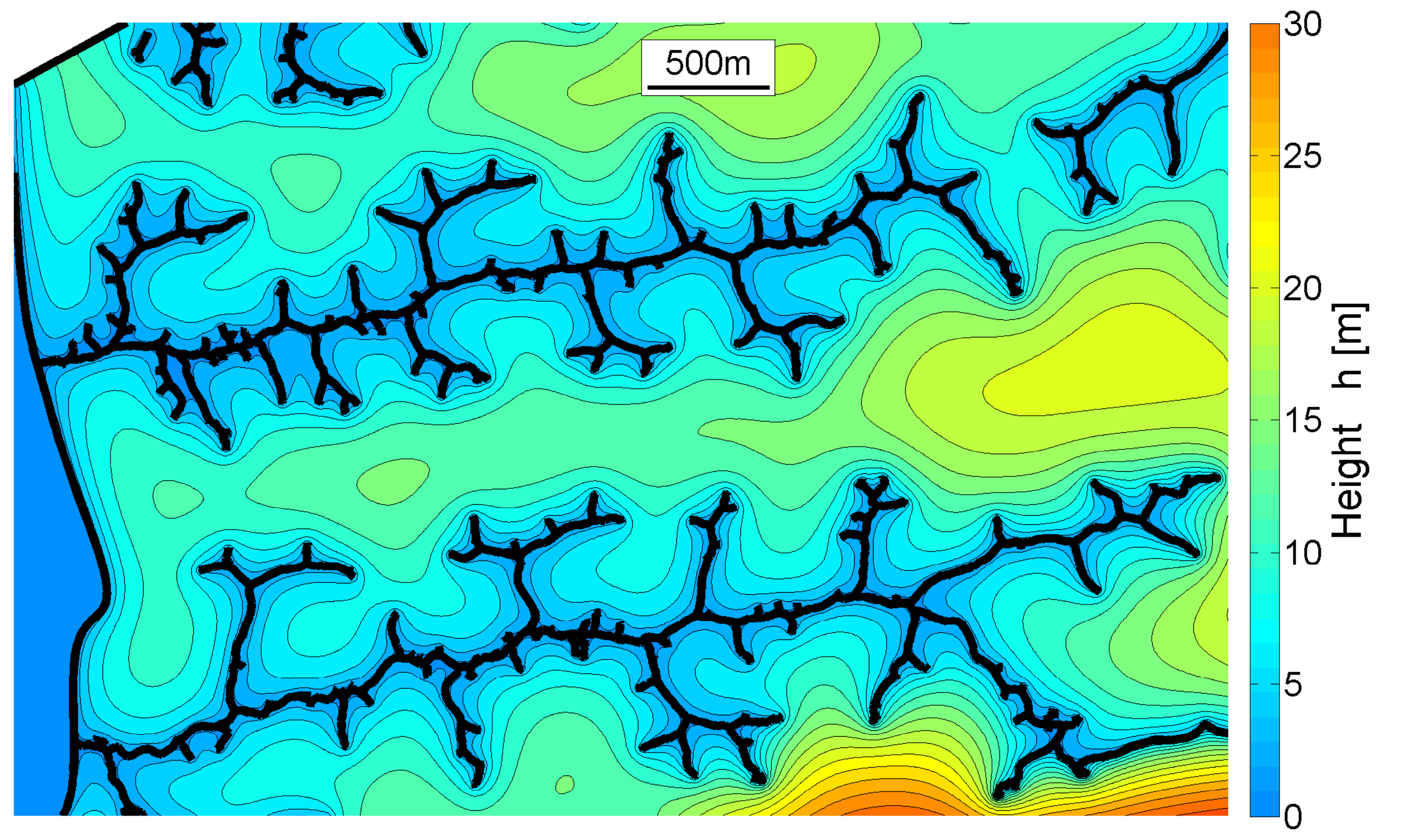}
  \caption{The height $h$ of the water table above a seepage network (black) in Bristol, Florida. The shape of the water table is calculated using Eq. \eqref{dupuit}, assuming the channel network is an absorbing boundary at $h=0$ and $P/\kappa =3.125\times10^{-3}$ \cite{09ALPSMMKR}.}
  \label{Florida}
\end{figure}

\subsection{Growth according to local symmetry}

We seek to determine if growth of a real stream network is consistent with the principle of local symmetry. We study a network of seepage valleys located near Bristol, Florida, on the Florida Panhandle \cite{09ALPSMMKR}. The network is presented in Fig. \ref{Florida}. In this network, groundwater flows through unconsolidated sand above the impermeable substratum, and into the streams \cite{95SBWS,09ALPSMMKR}. The flow is determined by the Poisson equation \eqref{dupuit}; thus the network grows in a Poisson field \cite{11PDALKR}.

We study the evolution of this network and check if the growth of the streams fulfills local symmetry. First, we set the boundary conditions; the change in elevation along the Florida network is small (the median slope $\sim10^{-2}$), we approximate the height of the channels above the impermeable layer as constant and choose $\phi(h=0)=0$. The outer boundaries are reflective, i.e. $\frac{\partial\phi}{\partial n}=0$, corresponding to a groundwater divide. The boundaries that close the domain are chosen arbitrarily. We calculate the Poisson field, Eq. \eqref{rdupuit}, and find for each channel head the coefficient $a_1$ in the expansion \eqref{expansion} that corresponds to the water flux entering the tip. Then, we remove a segment, $l_i$, from the tip of the $i$-th tributary and propagate it forward to its original length in five small steps (to reduce numerical error). The growth of each stream is characterized by two variables: its growth rate and the direction of its growth. We assume that the velocity of a stream is proportional to the magnitude of the gradient of the field, raised to a power $\eta$:
\begin{equation}
v\sim|\nabla\phi|^\eta\sim a_1^\eta.
\end{equation}
A similar growth law has been considered in Laplacian path models \cite{02CM,08GS,83HK,99SM,99WT}.
Thus, the length $l_i$ of each segment that we remove from a channel, and later add as it grows forward, is defined according to its relative velocity; $l_i\propto v_i/\langle v\rangle$, where $v_i\propto a_{1i}$ of the $i$-th channel and $\langle v\rangle=\frac{1}{n}\sum_i v_i$ is the mean velocity. We fix the total length removed from the network of $n$ tributaries to be $n$ meters.  Each channel then grows in a direction that fulfills local symmetry, i.e. in the direction for which $a_2$ vanishes. After we grow the network back to its original length, we study each of the tributaries separately, and measure the angle, $\beta$, between the real trajectory of the stream and the reconstructed trajectory. We perform this calculation for 255 channel heads in the Florida network. We obtain the mean of $\beta$ around zero with a standard deviation of $\sim7$. We find that a mean around $0$ of the angle error is consistent with a growth that fulfills local symmetry. However, some of the streams deviate significantly from their real growth direction, which may suggest that other factors account for their growth.

\begin{figure}
  \includegraphics[scale=1.5]{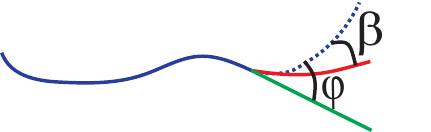}
  \caption{A retracted channel (blue) with two growth mechanisms: following local symmetry (red), and continuing in the tangent direction (green). The error angles $\beta$ and $\phi$ are calculated between a real trajectory of a channel (dashed blue) and a trajectory that follows local symmetry, and a real trajectory and the tangent, respectively. We find that the mean absolute value of the error angle measured for 255 channels in Florida is $\langle|\beta|\rangle=3.67$ for the local symmetry and $\langle|\varphi|\rangle=5.62$ for the tangent direction.}
  \label{angles}
\end{figure}

To evaluate the significance of the results, we suggest a null hypothesis in which the streams grow in the direction of the tangent regardless of the value of $a_2$. We obtain the direction of the tangent based on last $2$ grid points (approx. $2$ meters) of the channel trajectory after retraction. Then, we calculate the angle, $\varphi$, between the tangent direction and the real trajectory, as shown in Fig. \ref{angles}. We find that a growth according to the principle of local symmetry reduce on average the error angle by $50\%$ compare to a growth in the direction of the tangent, and therefore, improve the predication of a future growth trajectory of a channel.

\subsection{Growth law}

\begin{figure}[ht]
\centering
\subfloat[$\eta=0$]{
\includegraphics[scale=0.7]{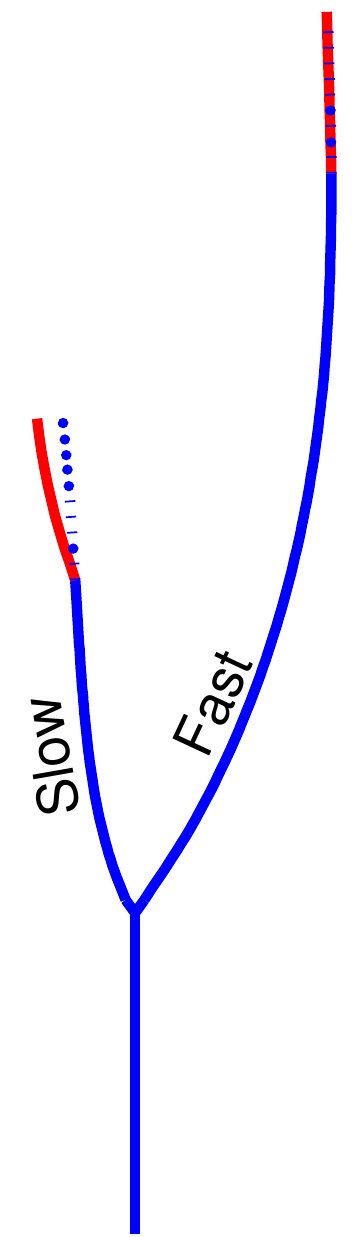}
\label{eta0}
}
\subfloat[$\eta=\eta_0$]{
\includegraphics[scale=0.7]{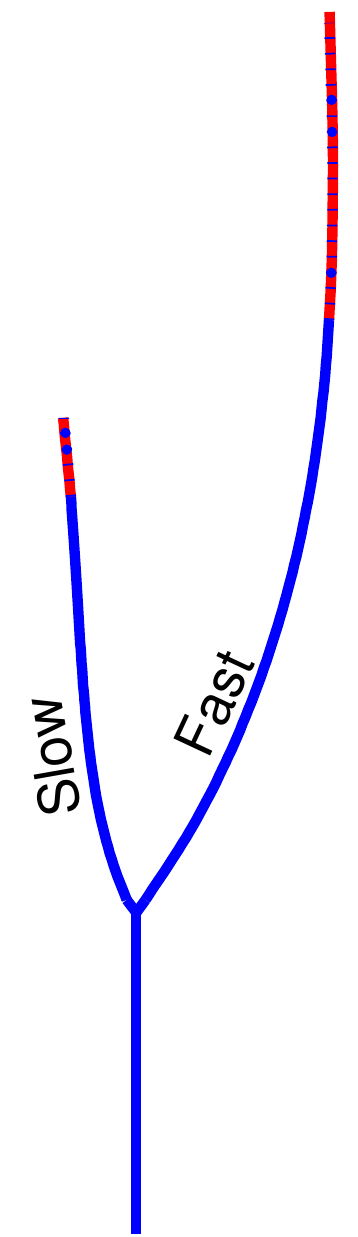}
\label{eta1}
}
\subfloat[$\eta>\eta_0$]{
\includegraphics[scale=0.7]{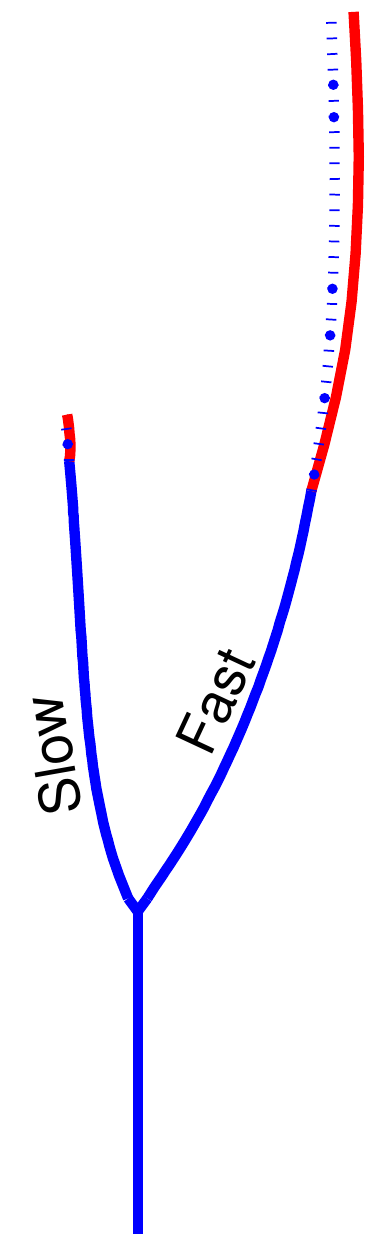}
\label{eta2}
}
\caption{A bifurcated channel (blue) is retracted with different growth exponents $\eta$; $\eta_0$ is the exponent that characterizes the original growth. The length of the retracted segment (dashed blue) is proportional to $a_1^\eta$. The faster stream has a bigger $a_1$. Thus, the segment that is removed from the faster stream becomes longer with respect to the slower stream as $\eta$ gets larger. The red curves are trajectories of channels grown forward. The deviation angle $\beta$ is measured between the real (dashed blue) and calculated (red) trajectories. In (a), deviations are characterized by growth away from the fast stream; in (c), the deviant growth avoids the slow streams.}
\label{velo}
\end{figure}

To understand the deviations between the real and the calculated path, we hypothesize that the deviant streams grew in a different environment than currently exists, e.g. the tributaries in the neighborhood of the stream were relatively undeveloped (or over-developed) when the studied stream reached its current location. To illustrate this idea, in Fig. \ref{velo} we show the trajectories of two streams with different velocities, and compare their evolution for different growth exponents $\eta$. One notices that for smaller $\eta$ the slower streams are more likely to deviate from their real trajectory, but for higher $\eta$ the faster streams change their course. Only when $\eta=\eta_0$ (the correct value of $\eta$), any errors will not be correlated to the velocity of the streams.

\begin{figure}[ht]
  \includegraphics[scale=0.5]{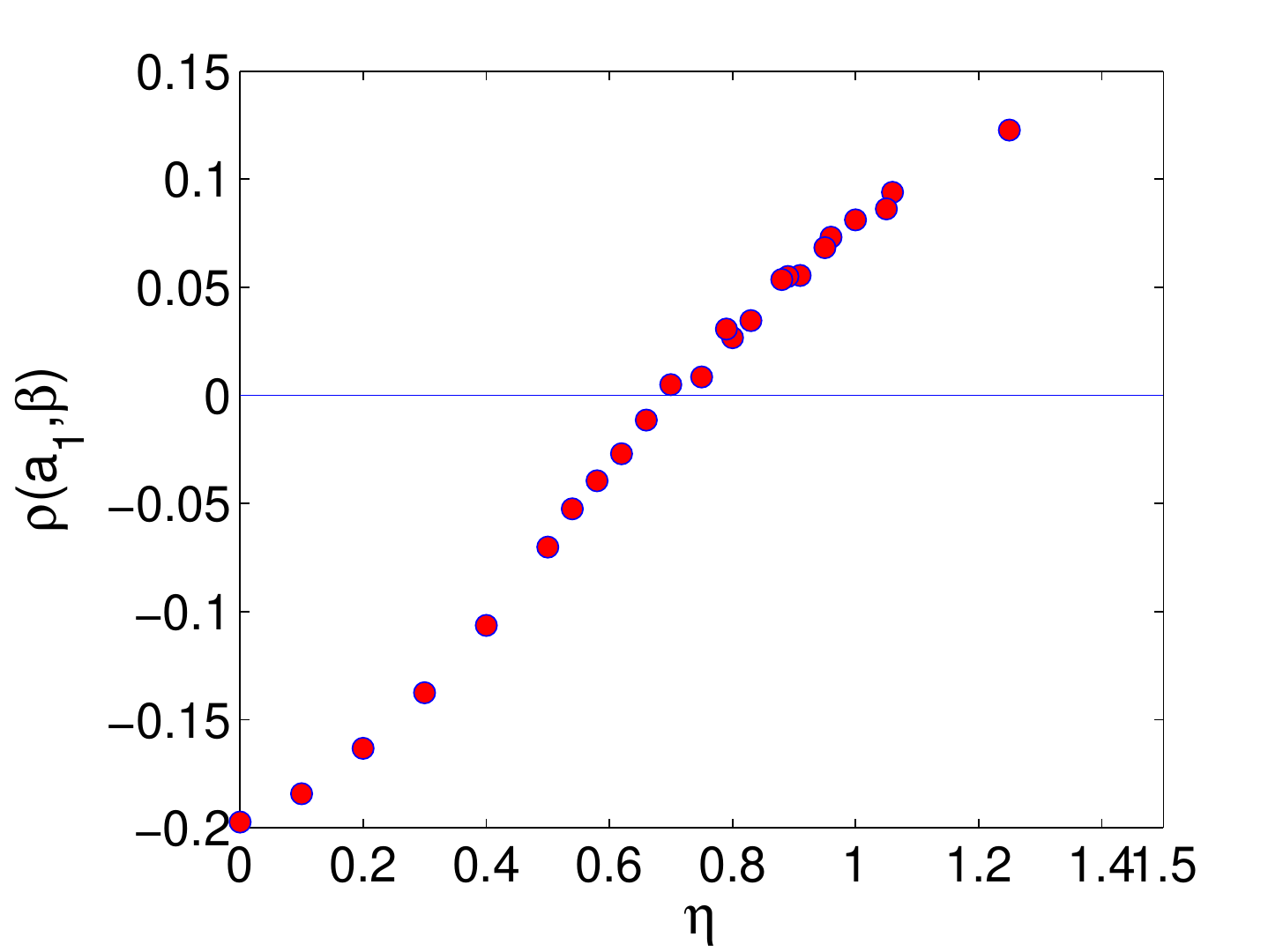}
  \caption{The Spearman rank correlation function $\rho(a_1,\beta)$ for different $\eta$ between $a_1$, the flux entering the tip of a channel, and $\beta$, the angle between the real trajectory of a stream and the calculated trajectory. The change in sign near $\eta=0.7$ suggests that the Florida network grew with an exponent $\eta\simeq0.7$.}
  \label{etas}
\end{figure}

Motivated by this reasoning, we study the correlation between the flux entering the tip, which we identify with $a_1$, and the angle $\beta$ for different values of $\eta$. Retracting the network with different $\eta$ creates different boundary conditions and influences the trajectory of the streams as they grow forward. For small $\eta<\eta_0$, and in particular when $\eta=0$, we remove the same segment size from each stream regardless of the magnitude of the flux entering the tip (Fig. \ref{eta0}). Thus, as we grow the network forward, the deviation from the real trajectory will be larger for the slower stream, with small $a_1$, since they try to avoid the faster streams that currently exist in their environment. Therefore, we expect that the deviation from the real trajectory will be larger for the slower stream. However, when $\eta>\eta_0$, the faster streams are retracted much further backward compared to slower streams, and they grow in a more developed network than the network that had existed when they had actually grown in the field. In this case, the faster streams will reveal a bigger error in their trajectory. For $\eta=\eta_0$, there is no correlation between the flux and $\beta$, which implies that $\eta_0$ is the best exponent for the growth. Fig. \ref{etas} shows that $\eta_0\simeq0.7$ for the Florida network.

\begin{figure}[ht]
  \includegraphics[scale=0.7]{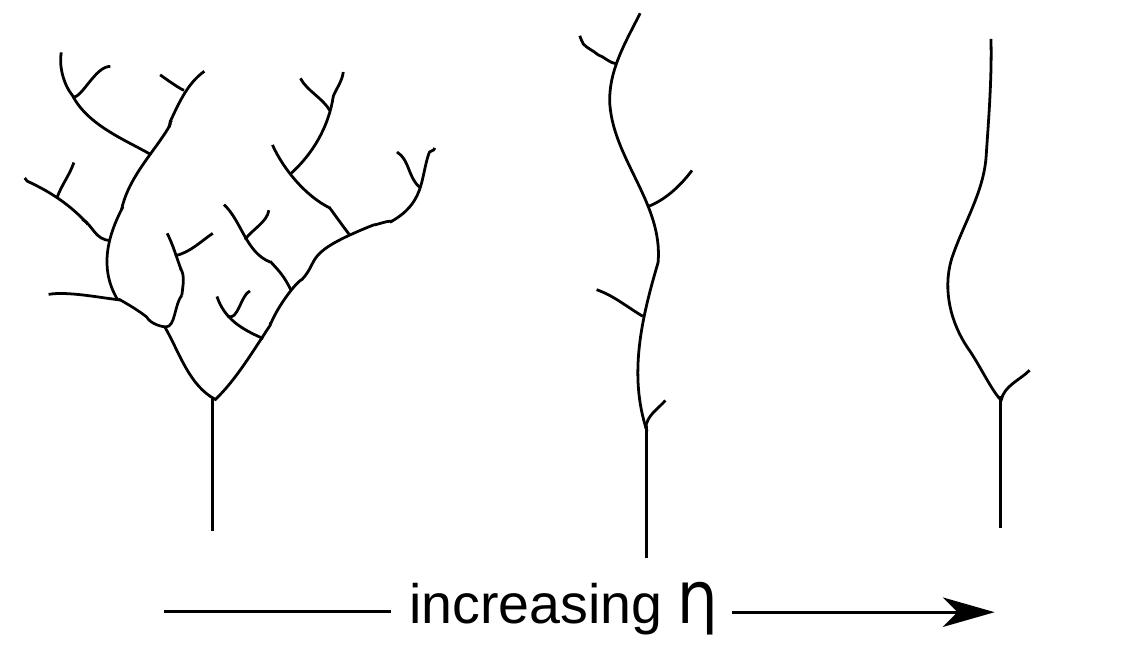}
  \caption{Illustration of the ramified structure for increasing growth exponent $\eta$.}
  \label{struc}
\end{figure}

The importance of the growth exponent $\eta$ is in the evolution of the network: negative $\eta$ will generate a stable network in which each perturbation, or small channel, will survive regardless of the water flux entering the tip. A positive $\eta$ results in an unstable structure in which a small difference in the velocity of two competing channels is amplified and may lead to a screening mechanism and the survival only of the faster channel \cite{08GS}. Fig. \ref{struc} contains a schematic representation of this concept. The small positive exponent found for the stream network in Florida indicates that this network is unstable. This conclusion is consistent with the prediction of a highly ramified network.

\section{Summary}
In summary, we offer a criterion for path selection of a stream in a diffusing field. We show that this criterion, which is based on the principle of local symmetry \cite{61BC,80CR}, predicts accurately the evolution of channels fed by groundwater. We suggest a method to infer the history of a real network by reconstructing it according the principle of local symmetry and evaluating errors for different growth laws. We parameterize the relationship between the water flux and the sediment transport with a single exponent and show that for the Florida network this growth exponent is about $0.7$. We envision that our methods may also be applied to other problems, such as the growth of hierarchical crack patterns \cite{88SM,05BPAAC,09CMP} and geological fault networks \cite{07DBS}, to provide a better understanding of their evolution.
\begin{acknowledgments}
We would like to thank The Nature Conservancy for access to the Apalachicola Bluffs and Ravines Preserve, and K. Flournoy and D. Printiss for guidance on the Preserve. This work was supported by Department of Energy Grant FG02-99ER15004.
\end{acknowledgments}

\appendix
\section{Propagation of a channel}
\begin{figure}[ht]
  \includegraphics[scale=0.5]{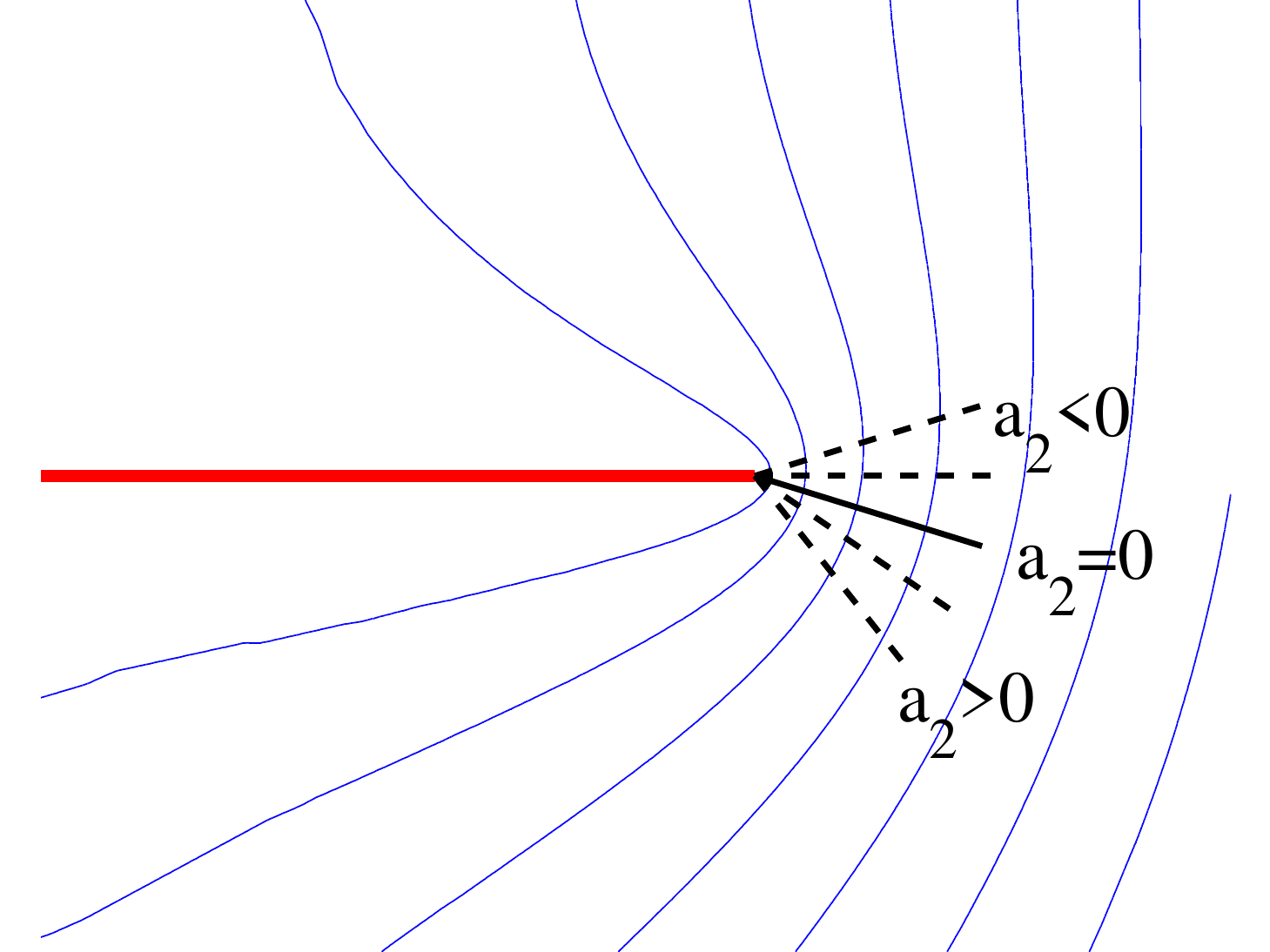}
  \caption{Several growth directions; A channel grows in a direction in which a$_2$ vanishes (solid black line).}
  \label{growthop}
\end{figure}

The evolution of a channel is defined by the field in its vicinity. In order to find the direction in which a channel tip evolves, we developed a numerical solver using Galerkin finite element discretization on a triangular grid \cite{KB00}, and solve the Poisson equation, Eq. \eqref{rdupuit} or the Laplace's equation Eq. \eqref{laplace}, with the described boundary conditions. Then we iteratively add a small segment to a stream in different directions, see Fig. \ref{growthop}. We obtain the field by using the numerical solver, and from the field of the vicinity of the studied stream, we find the coefficients of the expansion \eqref{expansion}. We accept the growth in the direction where the asymmetric coefficient becomes zero, $a_2\sim0$.

\section{Principle of local symmetry appears in Loewner growth.}
\begin{figure}[ht]
\center\includegraphics[scale=0.5]{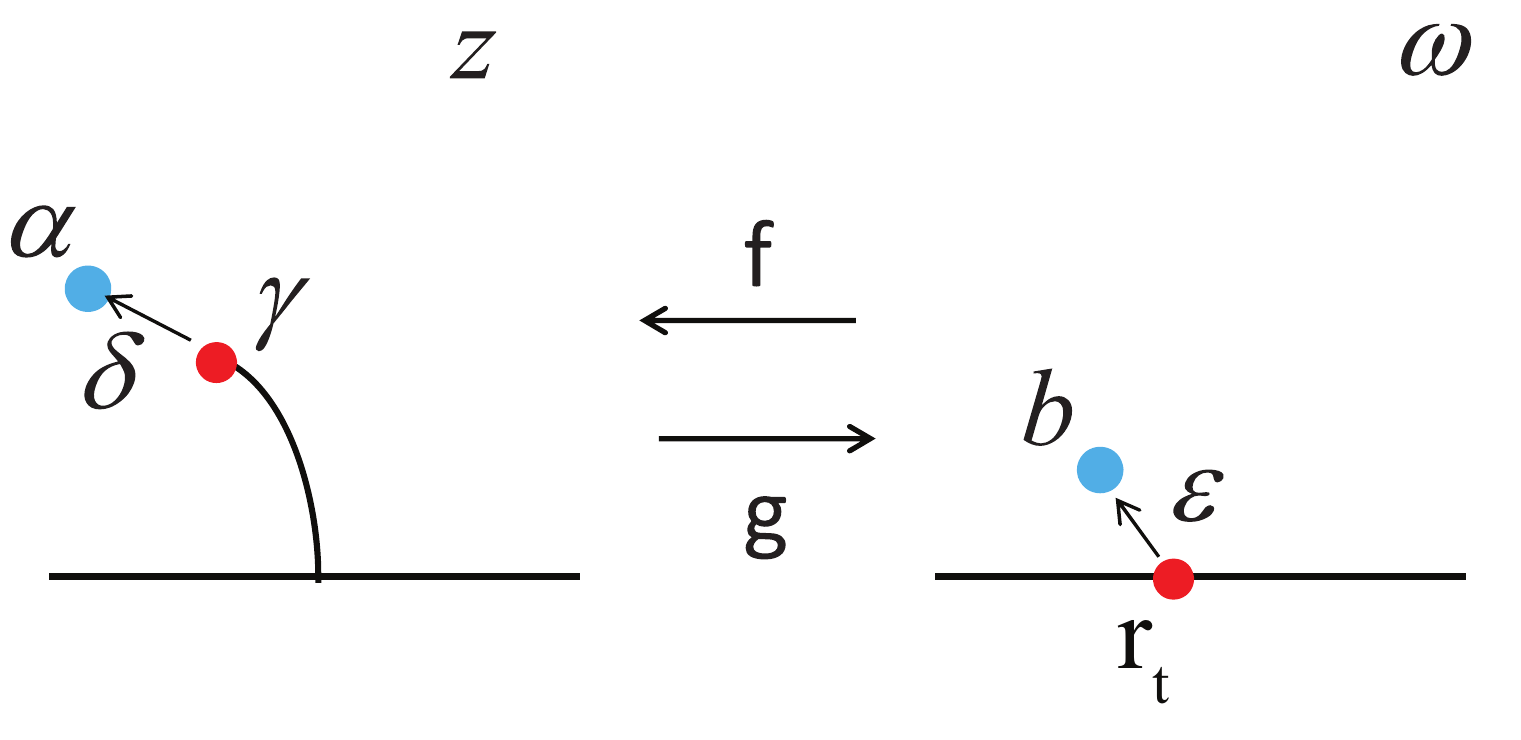}
\caption{An example of an arbitrary channel mapped into the real axis. The point $r_t$ indicates the position of the tip $\gamma$ on the real axis at time $t$.}\label{linear}
\end{figure}

The Loewner approach has been used to describe the evolution of curved path in the complex plane according to the gradient, or the streamlines, of the field in their neighborhood \cite{04GK,06BB,08GS}. Here, we show that the condition of a vanished asymmetric term $a_2 z$ in the expansion of the field near the tip,
\begin{equation}
 p(z) = a_1 z^{1/2} + a_2 z + ...
\end{equation}
is equivalent to the growth of the channel along the streamlines in successive moments of time. $p(z)$ is the complex expansion of the field near the tip, Eq. \ref{expansion}. The coefficients $a_i$, $i=1,2,..$ are real numbers, and $z=x+\imath y$ is a complex representation of the distance from the tip \cite{13PDSR}. We use a conformal mapping approach \cite{08GS}, in which the outside of the channel is mapped onto an empty half-plane (Fig. \ref{linear}) and the tip of the channel $\gamma$ mapped onto $r_t$. Next, $\alpha$ is a point in the neighbourhood of $\gamma$, mapped onto $b$, i.e.:

 \begin{equation}
 g(\gamma)=r_t
 \end{equation}
 \begin{equation}
 g(\alpha)=b
 \end{equation}
 \begin{equation}
 f(r_t)=\gamma
 \end{equation}
 \begin{equation}
 f(b)=\alpha
 \end{equation}

 Also, let us introduce
 \begin{equation}
\delta=\alpha-\gamma
 \end{equation}
 and
 \begin{equation}
 \epsilon=b-r_t
 \end{equation}
 as marked in the Fig. \ref{linear} . Our goal is to express the field around $\gamma$ as a function of $\delta$. The solution of the Laplace equation in the empty-half plane is simply $\Psi(\omega)=\omega$, where $\omega$ is a complex variable in the mapped plane. Thus, the field around the channel is given by $p(z)=g(z)$, i.e.
 \begin{equation}
 p(\alpha)-p(\gamma)=b-r_t=\epsilon
 \end{equation}
 or
 \begin{equation}
 p(\alpha)=p(\gamma)+\epsilon.
 \end{equation}
 To proceed further, we express $\epsilon$ as a function of $\delta$:
 \begin{equation}
\delta= f(b)-f(r_t) =1/2 f''(r_t) \epsilon^2 + 1/6 f'''(r_t) \epsilon^3+...
\label{star}
\end{equation}
Note that the term with $f'$ vanishes since the tip of the channel corresponds to the local maximum of $f$.
Next, we observe that $\delta$ vanishes when $\epsilon$ vanishes and for small $\delta$ the dependence is $\epsilon \sim \delta^{1/2}$. Thus we can look for $\epsilon(\delta)$ in the form of an expansion
\begin{equation}
 \epsilon(\delta) = a_1 \delta^{1/2} + a_2 \delta + ...
\end{equation}
Inserting this into \eqref{star} leads to
\begin{equation}
\delta = 1/2 f''(r_t) a_1^2 \delta +  f''(r_t) a_1 a_2 \delta^{3/2} + 1/6 f'''(r_t) a_1^3 \delta^{3/2} + ...
\end{equation}
Equating the coefficients at the same powers of $\delta$ we have
\begin{equation}
a_1= (f''(r_t)/2)^{-1/2} \label{c1}
\end{equation}
\begin{equation}
a_2 = -\frac{1}{6} a_1^2 \frac{f'''(r_t)}{f''(r_t)} = \frac{1}{3} \frac{f'''(r_t)}{f''(r_t)^2}.
\end{equation}
This leads us to the conclusion that $a_2$ vanishes when $f'''(z=r_t)$ vanishes and vice versa.
However, the condition of vanishing $f'''(z=r_t)$ is equivalent to the condition that the second derivative of the mapping  $f''(z=r_t)$ is at the extremum. Since, following Eq. \eqref{c1}, $|f''(z=r_t)|^{-1/2}$ determines the flux at the top of the channel at time $t$, the above condition guarantees that the flux is maximized and that growth is in the direction that fulfills local symmetry.

\bibliography{PLS}
\bibliographystyle{unsrt}

\end{document}